\PassOptionsToPackage{inline}{enumitem}
\documentclass[english]{lni}

\usepackage{todonotes}
\usepackage{csquotes}

\usepackage[inline]{enumitem}
\usepackage{calc}

\addto\extrasenglish{%
}

\usepackage[bookmarks=false]{hyperref}

\begin{document}

\title{From Requirements to Architecture: Semi-Automatically Generating Software Architectures}
\subtitle{PhD Research Proposal}
\date{\today}
\author[1]{Tobias Eisenreich}{tobias.eisenreich@tum.de}{0009-0004-7168-251X}
\affil[1]{Technische Universität München\\Chair of Software Engineering\\Bildungscampus 2\\74076 Heilbronn}
\booktitle[EMISA 2025]{15th International Workshop on Enterprise Modeling and Information Systems Architectures}
\yearofpublication{2025}

\maketitle

\begin{abstract}
To support junior and senior architects, I propose developing a new architecture creation method that leverages LLMs' evolving capabilities to support the architect. 
This method involves the architect's close collaboration with LLM-fueled tooling over the whole process. 
The architect is guided through Domain Model creation, Use Case specification, architectural decisions, and architecture evaluation. 
While the architect can take complete control of the process and the results, and use the tooling as a building set, they can follow the intended process for maximum tooling support.
The preliminary results suggest the feasibility of this process and indicate major time savings for the architect.
\end{abstract}


\section{Introduction}

Creating high-quality software architecture is hard due to complex domains and incomplete requirements.
New software architects lack experience and technical expertise, and experienced ones focus on the concepts and technologies they are used to~\cite{souza2019deriving}.
To avoid high maintenance costs from bad architectures, architects must explore broader possibilities and conduct a later evaluation.

I propose the development of supportive tooling for software architects leveraging the new advancements of LLMs in NLP.
This tooling should help novice architects to create high-quality architectures and nudge senior architects to explore new possibilities.

Colocated with ICSE'24, I discussed tool candidates with the research community~\cite{eisenreich2024requirements}:
\begin{enumerate*}[label=(\arabic*)]
\item Generation and refinement of domain models and use cases based on requirements;
\item generation and refinement of architecture candidates based on requirements, domain models, and use cases using established patterns; and
\item qualitative evaluation of architectures against the requirements.
\end{enumerate*}

I will evaluate these tools individually and in orchestration.
I believe this process can improve the quality of architecture while reducing the creation time in most projects.

\section{Background and Related Work}

Looking at the background, I will introduce research concerning software architecture and architecture evaluation methods.
The related work is brief, as not many papers have been published about the usage of AI in software architecture tooling.
Finally, I will describe my previous work and preliminary results.

\textcite{martin2017clean} wrote an overview of software architecture. 
He summarizes the learnings from multiple decades of experience in software engineering and architecture and draws a comprehensive image of software architecture. 
He describes and reasons about best practices, patterns, and frameworks. 
Many further works go into detail on these topics. 
A more concrete approach is \emph{Domain Driven Design}~\cite{evans2004domain}, which describes how domain modeling can be the basis for architecture creation. 
This approach is very popular in practice, and I take it as an inspiration for my work.
\textcite{patidar2015survey} gave an overview of software architecture evaluation methods. 
The Architecture Tradeoff Analysis Method (ATAM)~\cite{kazman2000atam} is probably the most known evaluation method. 
It helps to find sensitivities and tradeoffs in architectural design.
It targets multiple quality attributes but is tedious, involves intensive effort from all relevant stakeholders, and did not gain much traction in practice.

Due to the novelty of LLMs, related work about using them in software architecture is scarce. A Scopus search for \texttt{LLM AND \enquote{Software Architecture}} yielded 21 results, four of which are relevant to this work:
\textcite{jahic2024state} analyzed the state of practice on using LLMs in software architecture. 
However, they found mostly enthusiasm. Low quality, copyright issues, and the effort to check the generated solutions impede the usage of LLMs in software architecture practice.
\textcite{dhar2024llms} generated Architecture Decision Records (ADRs) with LLMs. They found that LLMs can generate relevant and accurate design decisions, albeit not yet with the same quality as humans.
\textcite{gustrowsky2024using} fine-tuned an LLM on AI-generated requirements on a small training set to predict one of three software architecture patterns for a single requirement. In their evaluation, seven of ten patterns were predicted correctly.
\textcite{gürtl2025design} used an LLM-based agent to aid first-term students in architectural decisions. Their results show that a chatbot cannot help students create substantially better architectures, but it can effectively teach them architecture fundamentals.

\subsection{Previous and In-Progress Own Work}\label{sec:wip}
I have outlined my vision of an assisted software architecture creation process before~\cite{eisenreich2024requirements}.
In the meantime, I have concretized the vision and worked on proofs-of-concept of the proposed toolings. 
I will give an overview of the progress in this section.

I am working on understanding architecture creation in the industry.
A student conducted several interviews with software architects from different contexts. 
We found that software architecture is a highly individual process, and the practice varies considerably between practitioners.

To support the architect, I am investigating the ability of LLMs to create Domain Models and Use Cases. 
My preliminary results are promising and suggest that an assisted creation of such models improves their quality while reducing the effort to create them.

A naive approach to generating architectures with an LLM does not produce quality architectures.
Asking an LLM for appropriate architectural styles or patterns works much better. 
One potential reason is the ability to use the RAG to describe architectural styles with known advantages and disadvantages, which helps the LLM to select appropriate ones.

Automating the architecture tradeoff analysis method~\cite{kazman2000atam} using LLMs works surprisingly well. 
While the LLMs do not find as many potential tradeoffs as the participants, the LLMs find a considerable overlap with the participants' responses.

\section{Research goal and objectives}

\begin{figure*}[tbp]
    \centering
    \includegraphics[width=\textwidth]{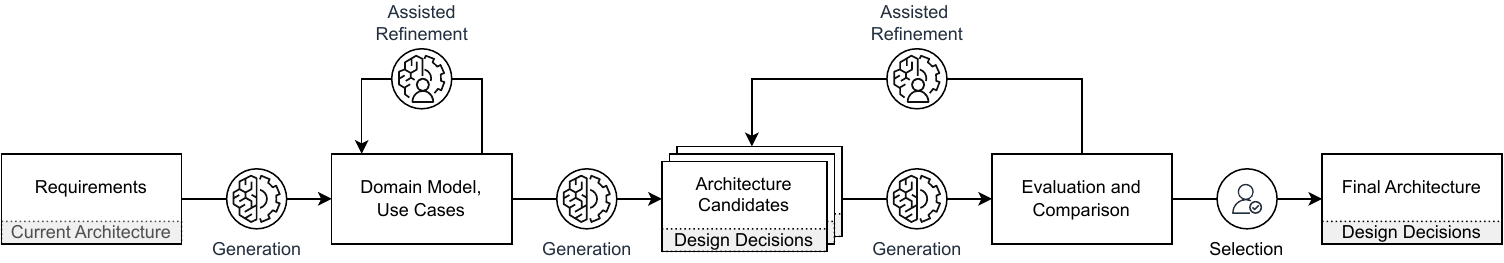}
    \caption{Tool-assisted architecture creation process.}
    \label{fig:process}
\end{figure*}

The research goal of the proposed work is to explore the new possibilities of assisting software architects using LLMs.
Looking at popular architecture creation processes like Domain Driven Design~\cite{evans2004domain}, I synthesized an exemplary, idealized process of  architecture creation.
I analyzed potentials for tool support during that process, as seen in \autoref{fig:process}.
I have started implementing these tools and obtained some preliminary results in their evaluation, as described in \autoref{sec:wip}.

From a technology perspective, I will choose prompt engineering over fine-tuning.
The obvious downside of this choice is forgoing some amount of response quality by not tuning the model to the use case.
There are a few advantages, though, that outweigh this downside:
First, there is a lack of labeled data; creating a high-quality labeled dataset for each tool to fine-tune a model would be a big effort.
Second, foundation models bring the knowledge for all possible application domains and can draw on it. Fine-tuning might favor the domains it has been trained with.
Lastly, prompt engineering techniques can easily be transferred between foundational models. The tooling can thus be tailored to the available hardware by using smaller models or be improved in the future by using newer, more capable models that are released after development time.

\section{Research Methodology}
As I plan to create one or multiple artifacts, the Design Science Research Method is a good fit for the scientific methodology~\cite{hevner2004design}.
Applying Design Science, I will evaluate both the individual tools and the whole idealized process with full tool support.

When evaluating the individual tools' quality, efficacy, and utility, I assume the biggest challenge is evaluating their quality.
The architecture creation process contains many design decisions for the architect, and in general it is hard to tell whether some are objectively better than others.
The quality of the tool output containing design decisions is not obvious and certainly not easily comparable.
Because of this issue, a meaningful quantitative evaluation becomes difficult, and I aim to evaluate with mixed methods instead.

This is what I am currently doing with the generation of domain models:
I use the developed tool to generate domain models from sets of requirements.
Building upon this, architects are tasked to create a domain model for the same requirements.
Half of the group can use the generated domain model as a basis and the other half creates the domain model from scratch.
Finally, a different group of architects is tasked to evaluate each domain model for its fit for the given requirements.
Using this methodology, I can evaluate both the plain tool output and how architects interact with it.

Another problem is acquiring real-world data to test the tooling.
I plan to draw on a collection of scientific model problems~\cite{shaw2025model} for the initial development and use real-world data gathered from cooperating companies only for later evaluation.

After all tools are evaluated separately, I want to evaluate the idealized process from \autoref{fig:process} as a whole.
For this, I will conduct a larger case study in cooperation with the industry.

\section{Threads}

I found three main threads to the doctoral project:
Other researchers explore using LLMs in software architecture, but I am confident I can contribute alongside them.
It may turn out that LLMs are inherently incapable of architecting software, but our experiments and advancements of foundational models suggest otherwise.
Evaluation is a primary concern for both LLM usage and design science, but it is difficult in this domain. I expect this to be an issue that accompanies me throughout the whole research process.

\section{Conclusion}
I propose research for a new tool-supported process that aids architects in their daily work. 
This process spans the whole creation of a new architecture and can be tailored by the architect. 
I believe this process can be a first step to improving the quality of industrial software architectures.

\printbibliography

\end{document}